\documentclass{iau}

\usepackage{amsmath}
\usepackage{graphicx}
\usepackage{multirow}

\def\HII{{H\,{\textsc{ii}}}}

\def\n2o2{{N2O2}}
\def\dd16{{N2S2 - N2H$\alpha$}}

\begin{document}

\lefttitle{Kathryn Grasha}
\righttitle{Wide Field Integral Spectroscopy in Nearby Galaxies with the TYPHOON Survey}

\jnlPage{1}{7}
\jnlDoiYr{2022}
\doival{10.1017/xxxxx}
\volno{373}

\aopheadtitle{Proceedings IAU Symposium}

\title{The Progressive Integral Step Method (PrISM) for Wide Field 3D Spectral Imaging of Nearby Galaxies: \\an Overview of the TYPHOON Survey}

\author{Kathryn Grasha}
\affiliation{Research School of Astronomy and Astrophysics, Australian National University, Canberra, ACT 2611, Australia; kathryn.grasha@anu.edu.au}
\affiliation{ARC Centre of Excellence for All Sky Astrophysics in 3 Dimensions (ASTRO 3D), Australia}
\affiliation{Visiting Fellow, Harvard-Smithsonian Center for Astrophysics, 60 Garden Street, Cambridge, MA USA}

\begin{abstract}
The TYPHOON program is producing an atlas of spectroscopic data cubes of 44 large-angular-sized galaxies with complete spatial coverage from 3650--9000~\AA. This survey provides an unparalleled opportunity to study variations in the interstellar medium (ISM) properties within individual \HII\ regions across the entire star-forming disks of nearby galaxies. This can provide key insights into the spatial distribution and \emph{resolved} properties of the ISM to understand how efficiently metals are mixed and redistributed across spirals and dwarf galaxies. In this Proceeding, we present early science results from six nearby spiral galaxies as part of the TYPHOON program from \citet{grasha22}. We use \texttt{HIIPhot} to identify the \HII\ regions within the galaxy based on the surface brightness of the H$\alpha$ emission line and measure variations of the \HII\ region oxygen abundance. In this initial work, we find that while the spiral pattern plays a role in organizing the ISM, it alone does not establish the relatively uniform azimuthal variations we observe across all the galaxies. Differences in the metal abundances are more likely driven by the strong correlations with the local physical conditions. We find a strong and positive correlation between the ionization parameter and the local abundances as measured by the relative metallicity offset $\Delta$(O/H), indicating a tight relationship between local physical conditions and their localized enrichment of the ISM. These variations can be explained by a combination of localized, star formation-driven self-enrichment and large-scale mixing-driven dilution due to the passing of spiral density waves.
\end{abstract}


\maketitle

\section{Introduction}

The `baryon cycle' --- the cycle of gas condensation, the resulting star formation and stellar feedback, and the re-injection of matter, energy, and momentum into the interstellar medium (ISM) --- is the engine for (secular) galaxy evolution and metal enrichment. This cycle of the star formation process is an inherently multi-phase and multi-scale process, dictating the assembly history and properties of galaxies by controling the flows of gas in and out of galaxies. Equally important is information on the internal gas dynamics and connecting the small-scale cycle of baryons with the larger galactic and circum-galactic scales. These processes drive the evolution of galaxies in the large-scale cosmological context, fueling the diversity of galaxies we see in the Universe today. At the core of these processes is the small-scale physics of star formation, which is intrinsically linked to and controlled by the feedback from massive stars \citep{hopkins13c, krumholz19}. A key unsolved challenge and area of modern astrophysics research, for both observations and theoretical models, is linking the observed global galaxy properties over cosmic time with the sub-parsec-scale physics of the star formation process.

Progress on the topic of galaxy evolution has rapidly moved into the domain of three–dimensional (3D) spectroscopy. Integral field spectroscopy (IFS) enables the spatially-resolved measurement of stellar populations, ionized gas, on-going star formation, star formation histories, kinematics, dust distributions, and chemical abundance maps. Numerous 3D Integral Field Units (IFUs) are now in operation and several more are in development for the next generation of telescopes. Moreover, a number of important IFU based galaxy surveys exist, such as SAURON \citep{dezeeuw02}, PINGS \citep{rosales-ortega10}, CALIFA \citep{sanchez12_califa}, SAMI \citep{croom12}, MaNGA \citep{bundy15}, SIGNALS \citep{rousseau-nepton19}, and PHANGS \citep{emsellem22}. Optical spectroscopy is a powerful tool to place the process of star formation in the context of its galactic host because it probes the ionized ISM that is intimately connected with young, massive stars. These surveys are making enormous progress in shedding light on the physical processes associated with star formation and variation with local conditions within local $z\sim0$ galaxies at the scales of individual star-forming regions \citep[e.g., ][]{kawamura09, sun20}. The ideal spatial scale to study is the size of individual star-forming products in galaxies (individual gas clouds, associations and clusters of young stellar objects, and discrete star-forming photoionized regions), which are typically a few tens of parsec in size \citep[e.g.,][]{grasha17, grasha18, ryon17, sun22}.

The production and build-up of heavy elements from stellar nucleosynthesis over cosmic time is one of the key physical properties for understanding galaxy evolution. The ISM oxygen abundance (i.e., the gas-phase metallicity) and distribution are built up by various physical processes throughout a galaxy's evolutionary history. The abundance of heavy elements is usually constrained by measuring the current gas phase oxygen abundance (metallicity) within \HII\ regions \citep{kennicutt96}. The gas-phase metallicity sets the balance between processes that enrich gas, such as star formation, and processes that dilute or remove metals, such as inflows of pristine gas from the intergalactic medium, galactic winds, and outflows and drive changes in fundamental local ISM properties (e.g., gas-to-dust ratio, CO-to-H2 conversion factor). Although these physical processes are often coupled in complicated ways, chemical evolution models have been able to place quantitative constraints on the key processes driving galaxy evolution \citep[e.g., ][]{kobayashi07, torrey19}. Thus \HII\ regions are strictly connected with the star formation process as they are powered by high mass ($>$8~M$_\odot$) stars and are fundamental in driving the evolution of galaxies. \HII\ regions are an important tool to investigate the star formation process and the properties of the ionized gas and the current chemical abundances of the gas from which stars are currently being produced. 

With the complete spatial coverage delivered by IFS, recent studies have ushered in a new era of characterizing oxygen abundance variations, where predominantly negative radial trends in metallicity, well established across large samples of galaxies, traces the inside-out growth of galaxy disks \citep{boissier99}. However, moving past simple 1D radial distributions and understanding the broader degree of chemical inhomogeneity in star-forming galaxies is key to making use of stellar metallicities within the field of Galactic archaeology and chemical tagging studies to assess accretion events and the formation history of galaxies \citep{buder22}.

The large FOV coverage of the TYPHOON data cubes (18' on one side, corresponding to deprojected galactocentric coverage of up to 35 kpc of the star-forming disk at the distances of the galaxies) provides us with an unparalleled sample to study variations in oxygen abundance in individual \HII\ regions across entire star-forming disks. The spatial coverage of 3650--9000~\AA\ enables constraints on the ISM pressure, density, and ionization parameter to provide an additional measure of the ISM conditions in these galaxies. Multiple diagnostics of the ISM conditions allows for the creation of simple chemical evolution model to understand the underlying physical processes governing observed variations in the chemical in-homogeneity of the ISM.

\section{Introduction to the TYPHOON Survey}

\begin{figure*}
\includegraphics[scale=0.62]{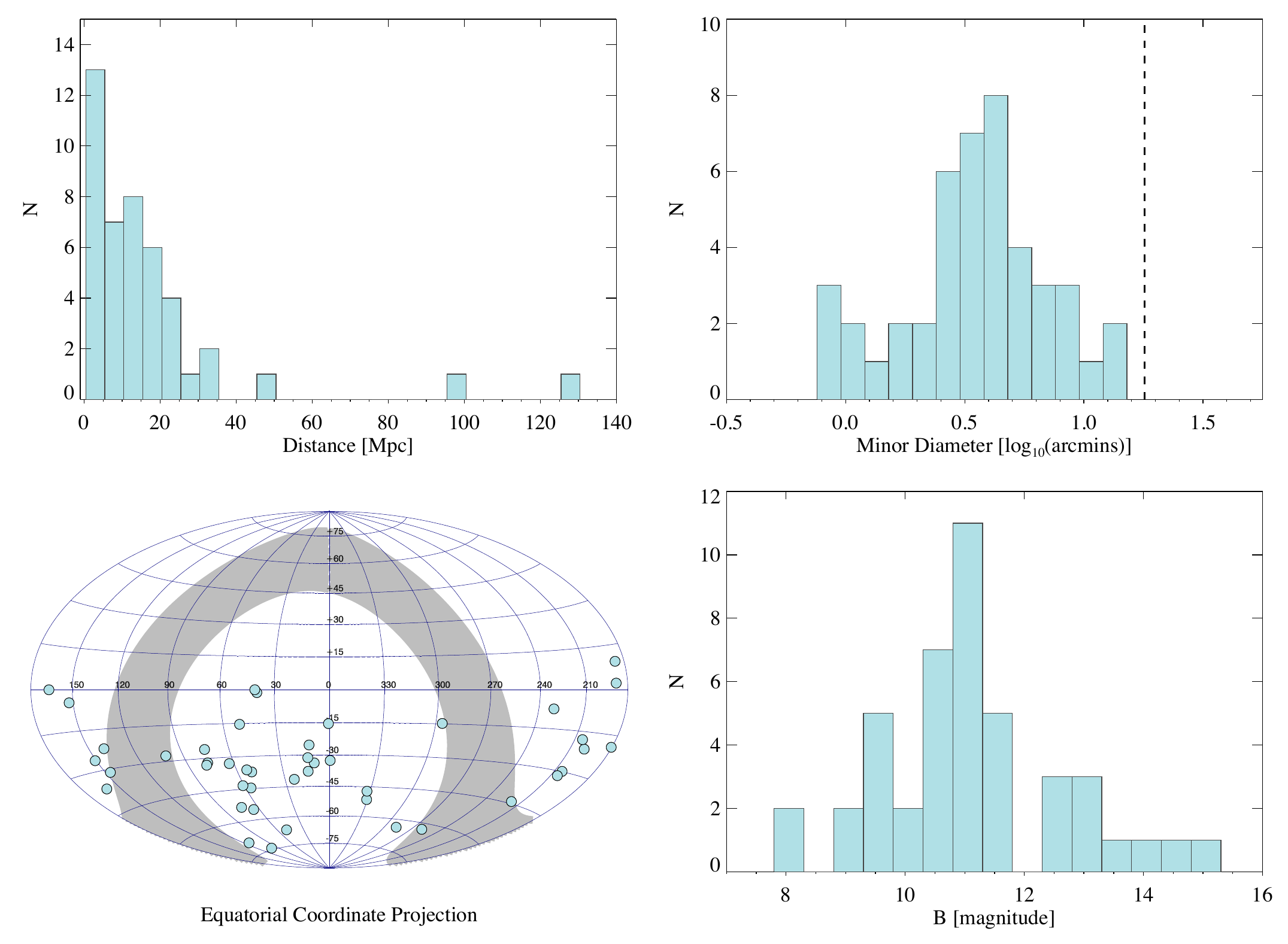}
\caption{Description of the PrISM/TYPHOON survey sample. The full sample is listed in Table~\ref{tab:1}.} 
\label{fig:survey}
\end{figure*}

We are producing an atlas of spectroscopic data cubes of 44 large-angular-sized galaxies (Figure~\ref{fig:survey} and Table~\ref{tab:1}). In the world of IFU applications, this is a largely overlooked (but still critically important) sample; to date they have been difficult to observe with classical IFUs as current IFU technology typically has small fields of view. The TYPHOON sample fills a niche in the 3D spectroscopy domain: current IFU surveys are optimized to observe galaxies at redshifts greater than $z\sim0.01$, and have spaxel sizes that resolve target galaxies at a physical scale of around 1~kpc. The simple method of PrISM is optimized (and only truly efficient) for bright large angular size galaxies at low redshift. This dataset of 44 local group galaxies within $\sim$20~Mpc and observable from the Southern Hemisphere will allow us to resolve at a median physical scale of 50 pc --- the physical scales of \HII\ regions \citep{azimlu11}. Early science results using TYPHOON galaxies have investigated the radial and azimuthal metallicity variations in select galaxies \citep{ho17,ho18}, the contribution of diffuse ionized gas to radial metallicity measurements \citep{poetrodjojo19}, and the starburst-active galactic nuclei mixing sequence \citep{dagostino18}. A survey description paper will be forthcoming (Seibert et al.). 

\subsection{Observational Techniques}

The TYPHOON Program\footnote{\url{https://typhoon.datacentral.org.au}} uses the Progressive Integral Step Method also known as ``step-and-stare'' or ``stepped-slit'' technique to construct 3D data of nearby galaxies. The PrISM data are the result of an observing methodology rather than a new instrument. We progressively step the slit of a conventional long-slit spectrograph across an object, building up a full data cube via a ``step and stare'' technique. We use PrISM with the Wide Field reimaging CCD (WFCCD) on the 2.5m du Pont telescope at Las Campanas Observatory. The WFCCD has a field of view of 25' with the 3D data cubes constructed using a custom long-slit (18'×1.65''; 0.5 arcminute$^2$) which is typically placed along the north–south direction and progressively scanned across each individual galaxy through stepping and staring. Each pointing position is integrated for 600~s before the slit was moved by one slit width in the east–west direction for the next integration. After each exposure, the slit is precisely stepped 1.65'' perpendicular to the spatial direction until the majority of the optical disk of each galaxy is covered. 

Images are taken at the beginning and end of each sequence of spectra, allowing us to align our observations from night to night. To minimize slit loss, observations are performed only when the seeing is less than the slit width of 1.65''. Spectrophotometric flux standards are observed each night.

\begin{table}
\begin{center}
\caption{The TYPHOON Sample.}\label{tab:1}
{\scriptsize\begin{tabular}{lllll||lllll}
\hline \hline
ID  & RA & Dec & Morph.   & D(Mpc) & ID  & RA & Dec & Morph.   & D(Mpc)  \\
\hline \hline
WLM	& 	00:01:58.16	&	-15:27:39.3	&	IB(s)m	&	1.04	&		NGC 2442	&	07:36:23.84	&	-69:31:51.0	&	SAB(s)bc 	&	21.01	\\
ESO 350--G 040	& 	00:37:41.11	& 	-33:42:58.8	&	S pec (Ring)  	&	130	&		NGC 2835	&	09:17:52.91	&	-22:21:16.8	&	SAB(rs)c     	&	10.36	\\
NGC 0253	&	00:47:33.12	&	-25:17:17.6	&	SAB(s)c	&	3.21	&		NGC 2997	&	09:45:38.70	&	-31:11:27.9	&	SA(s)c	&	9	\\
NGC 0289	&	00:52:42.36	&	-31:12:21.0	&	SAB(rs)bc	&	19.16	&		NGC 3109	&	10:03:06.88	&	-26:09:34.5	&	SB(s)m	&	1.33	\\
NGC 0300	&	00:54:53.48	&	-37:41:03.8	&	SA(s)d	&	1.94	&		Sextans A	&	10:11:00.80	&	-04:41:34.0	&	IBm	&	1.43	\\
NGC 0625	&	01:35:04.63	&	-41:26:10.3	&	SB(s)	&	3.78	&		NGC 3354	&	10:43:02.90	&	-36:21:44.6	&	S: pec         	&	49	\\
NGC 1015	&	02:38:11.56	&	-01:19:07.3	&	SB(r)a	&	35.1	&		NGC 3521	&	11:05:48.58	&	-00:02:09.1	&	SAB(rs)bc	&	11.39	\\
NGC 1068	&	02:42:40.71	&	-00:00:47.8	&	(R)SA(rs)b	&	10.58		&	ARP 244	&	12:01:53.30	&	-18:52:37.0	&	HII	&	20.94	\\
NGC 1313	&	03:18:16.05	&	-66:29:53.7	&	SB(s)d         	&	4.13	&		NGC 4424	&	12:27:11.61	&	+09:25:14.4	&	SB(s)a	&	14.11	\\
NGC 1309	&	03:22:06.56	&	-15:24:00.2	&	SA(s)bc	&	28.13	&		NGC 4536	&	12:34:27.05	&	+02:11:17.3	&	SAB(rs)bc	&	15.04	\\
NGC 1316	&	03:22:41.72	&	-37:12:29.6	&	SAB(s)	&	19.18	&		NGC 5068	&	13:18:54.81	&	-21:02:20.8	&	SB(s)d	&	5.88	\\
NGC 1365	&	03:33:36.37	&	-36:08:25.4	&	SBb(s)b	&	16.56	&		NGC 5236	&	13:37:00.95	&	-29:51:55.5	&	SAB(s)c	&	6.51	\\
NGC 1448	&	03:44:31.92	&	-44:38:41.4	&	SAcd	&	16.43	&		NGC 5247	&	13:38:03.04	&	-17:53:02.5	&	SA(s)bc	&	9.35	\\
NGC 1512	&	04:03:54.28	&	-43:20:55.9	&	SB(r)ab        	&	12.49	&		NGC 5253	&	13:39:55.96	&	-31:38:24.4	&	Im pec	&	3.65	\\
NGC 1532	&	04:12:04.33	&	-32:52:27.2	&	SB(s)b 	&	17.89	&		NGC 5643	&	14:32:40.74	&	-44:10:27.9	&	SAB(rs)c    	&	11.44	\\
NGC 1566	&	04:20:00.42	&	-54:56:16.1	&	SAB(rs)bc	&	9.95	&		NGC 5917	&	15:21:32.57	&	-07:22:37.8	&	Sb 	&	31	\\
NGC 1705	&	04:54:13.50	&	-53:21:39.8	&	SA0- pec	&	5.32	&		NGC 6300	&	17:16:59.47	&	-62:49:14.0	&	SB(rs)b       	&	12.26	\\
NGC 1744	&	04:59:57.80	&	-26:01:20.0	&	SB(s)d	&	9.99	&		NGC 6744	&	19:09:46.10	&	-63:51:27.1	&	SAB(r)bc       	&	7.24	\\
NGC 1800	&	05:06:25.72	&	-31:57:15.2	&	IB(s)m         	&	7.71	&		NGC 6822	&	19:44:57.74	&	-14:48:12.4	&	IB(s)m	&	0.52	\\
UGCA 106	&	05:11:59.32	&	-32:58:21.4	&	SAB(s)m	&	10.2	&		IC 5152	&	22:02:41.51	&	-51:17:47.2	&	IA(s)m	&	2.16	\\
ESO 034--IG 011	& 	06:43:06.10	&	 -74:13:35.0	&	(S0-) + Ring	&	99		&	NGC 7213	&	22:09:16.31	&	-47:09:59.8	&	SA(s)	&	22	\\
NGC 2280	&	06:44:49.11	&	-27:38:19.0	&	SA(s)cd       	&	22.9	&		NGC 7793	&	23:57:49.83	&	-32:35:27.7	&	SA(s)d  	&	3.83	\\
\hline \hline
\end{tabular}} 
\end{center}
\vspace{-10pt}
\tabnote{Coordinates (RA/Dec), morphological type, and distances (Mpc) are taken from the NASA/IPAC Extragalactic Database.}
\end{table}

\subsection{Data Reduction and Cube Construction}

\begin{figure*}
\includegraphics[scale=0.43]{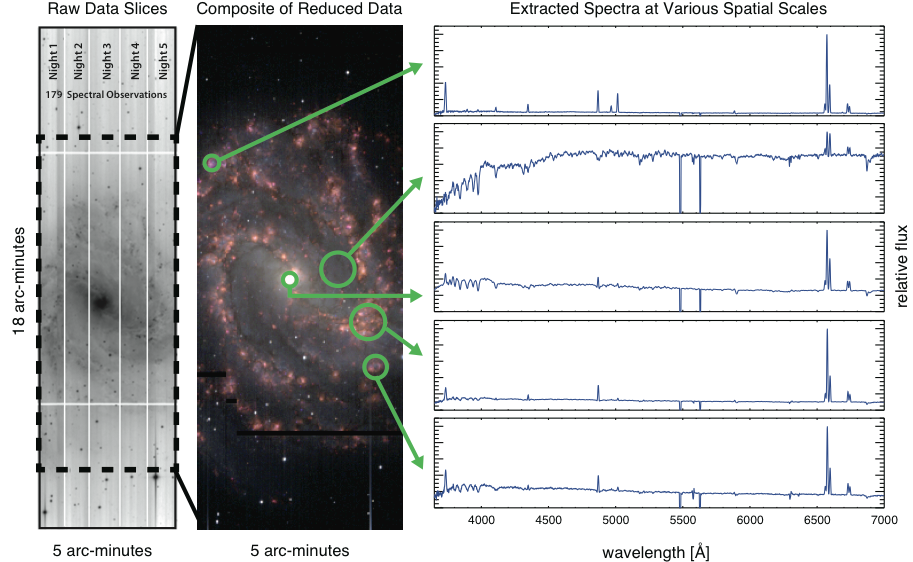}
\caption{Schematic visualization of NGC~5236 and extracted spectra at various spatial scales and locations.} 
\label{fig:m83}
\end{figure*}

Each individual frame is reduced using our custom-built pipeline. The spectra are bias-subtracted and flat-fielded using calibration frames taken each night. Exposures of a HeNeAr arc lamp are used to calculate a 2d wavelength solution for each spectrum, and each frame is rectified using a spatial solution generated with a Hartmann mask. Each frame thus results in a single 1.65'' $\times$ 18' ``slice'' of the full datacube. We bin each slice along the spatial direction of the slit from the native 0.484'' pixel size to 1.65''. The wavelength calibration has a typical root-mean square value 0.05~\AA\ for the entire data set. Flux calibration is accurate to 2\% at the spaxel scale over the range of 4500–7500~\AA. WCS solutions are individually obtained for single-night sections of the datacube using the astrometry.net software based on sources selected from GAIA DR2 or DR3. The reduced long-slit 2D spectra are tiled together to form the 3D data cube and we achieve astrometric solutions that are accurate to within 1 spaxel (offsets$<$1.65''). The final reduced data cube of each galaxy covers a wavelength range of 3650--9000~\AA\ at a spectral resolution of R$\sim$800, with spectral and spatial samplings of 1.5~\AA\ and 1.65'', respectively (Figure~\ref{fig:m83}). 

\subsection{Datacube Processing -- LZIFU and HIIphot}
The emission-line fluxes are measured using LZIFU \citep{ho16}. The emission lines are fitted as a single Gaussian component using the Levenberg–Marquardt least-squares method and the stellar continuum is modeled and subtracted for each spaxel. Further details on the LZIFU fitting procedure can be found in \citet{ho17}. 

\begin{figure*}
\includegraphics[scale=0.35]{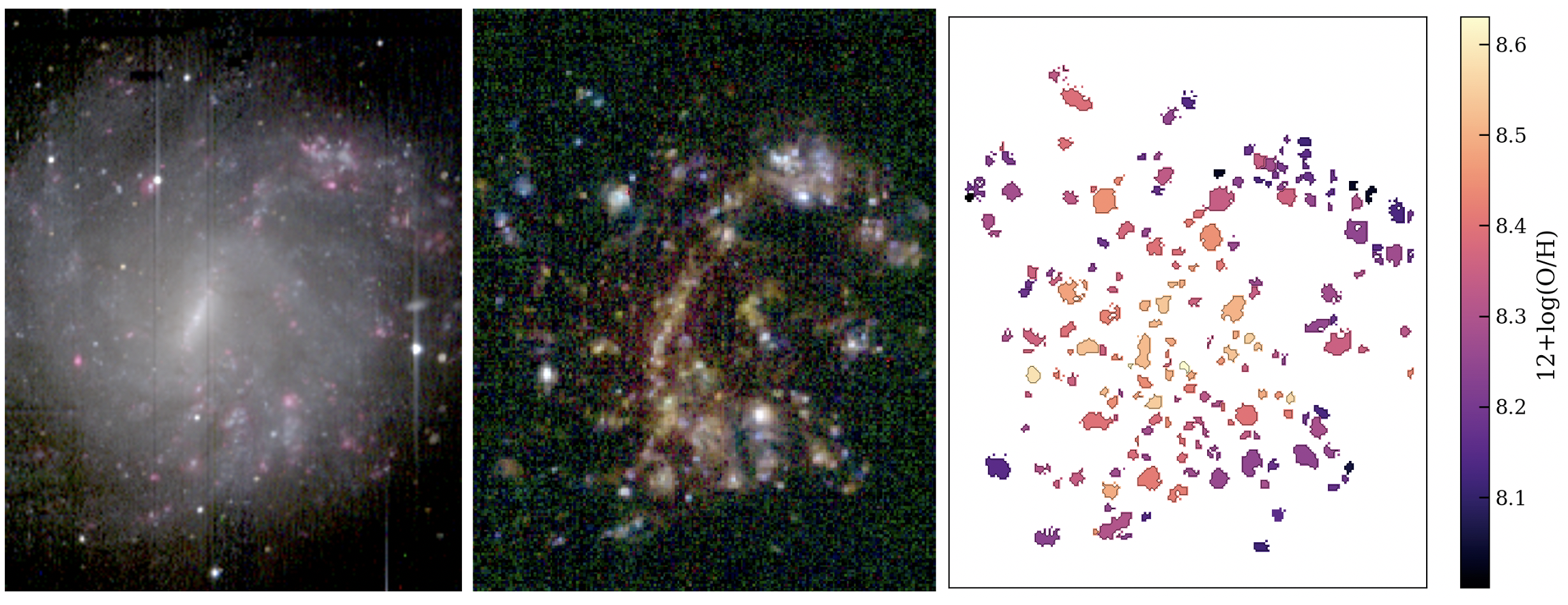}
\caption{A composite image shows the stellar continuua of NGC 5068 (left), the H$\beta$, [OIII]$\lambda$5007, and H$\alpha$ composite emission line image (middle), and the extracted HII regions colored by their gas-phase metallicity 12+log(O/H) \citep[right,][]{grasha22}.} 
\label{fig:ngc5068}
\end{figure*}

We use continuum subtracted H$\alpha$ emission line maps to identify the HII regions using HIIphot \citep{thilker00}. HIIphot morphologically isolates HII regions from background diffuse ionized gas (DIG), and provides a 2D mask that identifies separate regions within the map. Figure~\ref{fig:ngc5068} displays the \HII\ region catalog for NGC~5068 and further details can be found in \citet{grasha22}. The \HII\ region catalog paper will be forthcoming (Grasha et al.).

\section{Early Science Demonstration of \citet{grasha22}}

In a study of the first six spiral galaxies of TYPHOON, \citet{grasha22} found `typical' negative radial metallicity gradients but also a flattening in the gradient in several of the galaxies at large galactocentric radius, suggesting that we have sufficient resolution and sensitivity to detect the metallicity plateau, set by direct accretion of gas from the circumgalactic medium. The measured metallicity floor informs on the chemical enrichment of the galaxy and the outflow of metals to the outer regions of the disk from the enriched inner region of the galaxy disk through radial flows. 

\begin{figure*}
\includegraphics[scale=0.45]{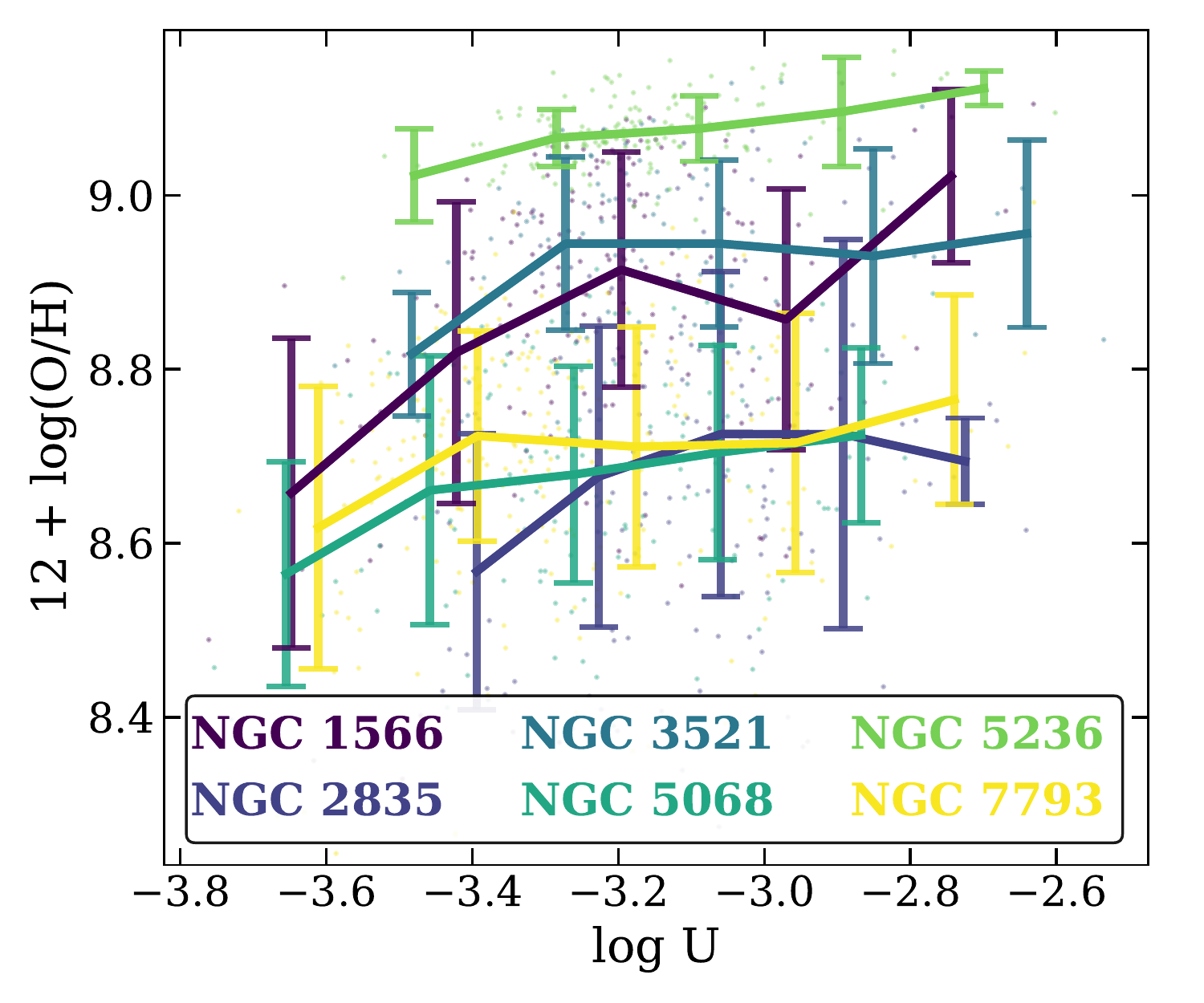}
\includegraphics[scale=0.45]{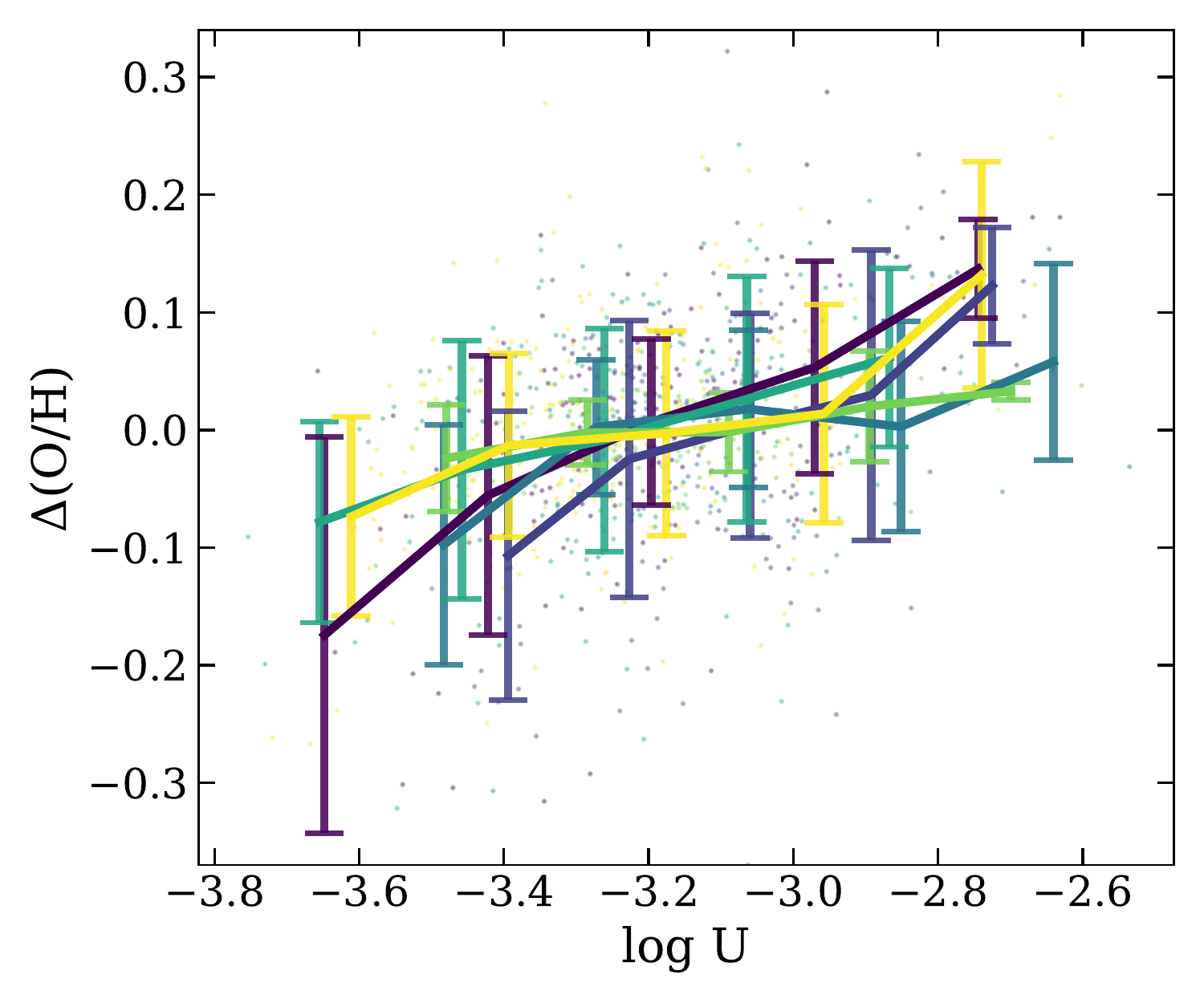}
\caption{Left: 12 + log(O/H) metallicity of each \HII\ region vs. the ionization parameter $U$ colored by each individual galaxy. Right: the metallicity offset from the linear radial gradient $\Delta$(O/H) vs. the ionization parameter $U$. There is a significant positive relationship between the residual metallicity $\Delta$(O/H) and the ionization parameter $U$, where \HII\ regions with higher metallicity offsets are association with higher ionization parameters.} 
\label{fig:g22}
\end{figure*}

We additionally find a lack of azimuthal variations in the oxygen abundance in the spirals in the survey. While the spiral pattern plays a role in organizing the ISM, it alone does not establish the relatively uniform azimuthal variations we observe. Differences in the metal abundances are more likely driven by the strong correlations with the local physical conditions. We find a strong and positive correlation between the ionization parameter $U$ and the local abundances as measured by the relative metallicity offset $\Delta$(O/H), indicating a tight relationship between local physical conditions and their localized enrichment of the ISM (Figure~\ref{fig:g22}). This is in agreement with expectations from a combination of localized, star formation-driven self-enrichment and large-scale mixing-driven dilution due to the passing of spiral density waves \citep{ho17}.

\acknowledgements
I gratefully acknowledge the Australian Research Council as the recipient of a Discovery Early Career Researcher Award (DECRA) Fellowship (project DE220100766) and the ARC Centre of Excellence for All Sky Astrophysics in 3 Dimensions (ASTRO 3D; project CE170100013). I also acknowledge the TYPHOON team for their assistance in writing of this paper: Barry Madore, Andrew Battisti, Rachael Beaton, Brent Groves, Lisa Kewley, Jeff Rich, and Mark Seibert.




\end{document}